\documentclass[aps,prl,reprint,amsfonts,amssymb,amsmath,nofootinbib]{revtex4-1}
\usepackage[english]{babel}
\usepackage{graphicx}
\usepackage{mathptmx}
\usepackage[utf8]{inputenc}
\usepackage[T1]{fontenc}
\usepackage{xcolor}
\usepackage{hyperref}
\usepackage{bm}
\hypersetup{final,colorlinks,linkcolor={red!50!black},citecolor={blue!50!black},urlcolor={blue!80!black}}
\graphicspath{ {figures/} }

\newcommand\mrm[1]{\mathrm{#1}}

\makeatletter
\newcommand{\closure}[2][3]{%
{}\mkern#1mu\overline{\mkern-#1mu#2}}
\newcommand*{\textoverline}[1]{\ensuremath{\closure[1]{\hbox{#1}}\m@th}}

\makeatother

\begin{document}
\title{Controlling rotationally-resolved two-dimensional infrared spectra with polarization}
\date{\today}
\author{Grzegorz Kowzan}
\email[Electronic address: ]{gkowzan@umk.pl}
\affiliation{Department of Chemistry, Stony Brook University, Stony Brook, NY 11790-3400, USA}
\affiliation{Institute of Physics, Faculty of Physics, Astronomy and Informatics, Nicolaus Copernicus University in Toru\'n, ul. Grudzi\k{a}dzka 5, 87-100 Toru\'n, Poland}
\author{Thomas K. Allison}
\affiliation{Department of Chemistry, Stony Brook University, Stony Brook, NY 11790-3400, USA}
\affiliation{Department of Physics and Astronomy, Stony Brook University, Stony Brook, NY 11790-3400, USA}

\begin{abstract}
Recent advancements in infrared frequency combs will enable facile recording of coherent two-dimensional infrared spectra of gas-phase molecules with rotational resolution (RR2DIR).
We describe how RR2DIR spectra can be controlled with polarization to enhance or suppress certain features of the spectrum, with new polarization conditions unique to freely rotating molecules and absent in the condensed-phase.
With the polarization control methods described here, RR2DIR spectroscopy can be a powerful tool for studying complex gas mixtures of polyatomic molecules.
\end{abstract}
\maketitle

Technological advancements in spectroscopy have been key drivers of advancements in both fundamental physics and myriad areas of applied science.
In the current era, many new developments in spectroscopy are being driven by rapid progress in optical frequency combs.
For example, frequency combs are now used in precision metrology~\cite{Hall2001,Ye2005}, searches for new physics beyond the standard model~\cite{Kennedy2020}, discovery of extrasolar planets~\cite{Steinmetz2008, Suh2018a}, advanced spectroscopy and imaging in the research laboratory~\cite{Weichman2019, Coddington2016, Cingoz2012}, and breath analysis in clinical settings~\cite{Liang2022}, to name just a few applications.

Most applications of frequency comb technology so far have used combs for \emph{linear} spectroscopy, in which the light-matter interaction is well-described by first-order perturbation theory, however several researchers have begun to explore the application of frequency combs to \emph{nonlinear} spectroscopy. 
For example, Allison and co-workers have used frequency comb techniques to enhance ultrafast transient absorption spectroscopy (a third-order response) in dilute gasses with detection limits approximately 4 orders of magnitude lower than conventional methods~\cite{Reber2016, Silfies2021}.
Allison also framed the more general coherent 2D spectroscopy in terms of wave mixing of multiple frequency combs and described methods for cavity-enhancing 2D spectroscopy signals~\cite{Allison2017}.
Lomsadze and Cundiff have demonstrated high-resolution coherent multidimensional spectroscopy in optically thick Rubidium vapors with multiple frequency combs~\cite{Lomsadze2018a}.
Meanwhile, others have pursued a hybrid approach performing (incoherent) double-resonance spectroscopy using a combination of a near-infrared frequency comb and an intense CW laser to saturate particular transitions~\cite{Foltynowicz2021}.

This previous nonlinear spectroscopy work has used combs in the near infrared and visible, where frequency comb technology is relatively mature.
Recent work on comb development has seen spectacular successes in the development of high-power and broadband frequency combs in the mid-infrared and long-wave infrared ($\lambda = 3$--$20$ $\mu$m) using a variety of platforms~\cite{Catanese2020,Xing2021,Nakamura2022,Ru2021,Gaida2018b,Lesko2021}.
In particular, the combination of high power and broad spectral coverage in the molecular fingerprint region ($\lambda = 3$--$20$ $\mu$m) inspires the idea of performing two-dimensional infrared spectroscopy on dilute gasses with resolution of rotational eigenstates.
Sensitive, high-resolution multidimensional spectroscopy in the fingerprint region could enable analysis of complex mixtures of polyatomic gases with unprecedented specificity.
Such mixtures commonly occur in human breath, flames and in detection of explosives and narcotics, but their linear spectra are highly congested and difficult to interpret.
The new spectroscopic capabilities would also benefit fundamental chemical physics studies on problems such as intramolecular vibrational redistribution and collisional processes in gasses~\cite{Mandal2018}.
With a large information density of rotationally resolved 2DIR (RR2DIR) spectra, there are likely many unforeseen applications as well. 
However, since such spectra have not been reported before, there is much work to do even to understand the fundamentals of RR2DIR spectra.

In this Letter we describe the control of RR2DIR spectra with polarization. 
Our work builds on previous work studying state-resolved four-wave mixing in gasses~\cite{Williams1994,Williams1994a,Williams1995,Williams1996,Williams1997,Wasserman1998,Bracamonte2003,Murdock2009,Wells2018}, but is more general in several ways.
First of all we consider the full third-order response accessible with three independent ultrashort pulses including rotationally coherent pathways (Feynman diagrams with rotational coherence during the waiting time between the second and third pulses) and we consider the effect of polarization on the complete response, as opposed to subsets of third-order pathways associated with specific experimental configurations.
We discover that inclusion of rotationally coherent pathways has important effects on the RR2DIR spectrum, including the inability to eliminate phase twist along the diagonal in absorptive 2DIR spectra.
Second, we discover several new polarization conditions for suppressing certain branches of the RR2DIR spectra.
These polarization conditions are distinct from those commonly used in condensed-phase 2DIR spectra and those previously considered in gas-phase four-wave mixing.
Finally, we illustrate the power of these techniques with an example regarding the separation of trace isotopes of methyl chloride.
In this short Letter, we focus on these novel features.
Our theory, described more fully in a longer companion paper~\cite{Kowzan2022} with an accompanying software package~\cite{Kowzan2022rotsim2d}, is completely general and can be used to simulate RR2DIR spectra under a range of conditions.

We consider the case of three ultrashort laser pulses interacting with a gas sample.
The third-order polarization is related to the incident field by:
\begin{multline}
  \label{eq:1}
  \hat{\epsilon}_{4}\cdot\vec{P}^{(3)}(\vec{r},t) = \iiint_{0}^{\infty}
  \mrm{d}t_{3}\,\mrm{d}t_{2}\,\mrm{d}t_{1} \;
  \hat{\epsilon}_{4}\cdot\mathbf{R}(t_{3}, t_{2}, t_{1}) : \big[\vec{E}(\vec{r},t-t_{3})\\ \vec{E}(\vec{r},t-t_{3}-t_{2}) \vec{E}(\vec{r},t-t_{3}-t_{2}-t_{1})\big],
\end{multline}
where $t_{i}$ are delays between interactions, $\mathbf{R}(t_{3}, t_{2}, t_{1})$ is the third order nonlinear response function, which is a fourth rank tensor, and ``$:$'' denotes tensor contraction---here, threefold contraction with electric field terms.
The unit vector $\hat{\epsilon}_{4}$ specifies the polarization detection axis.
We consider time-ordered excitation by three ultrashort optical pulses.
The direction and frequency of emitted third-order field is specified by $\vec{k}_{s}$, $\omega_{s}$:
\begin{align}
  \label{eq:2}
   \vec{k}_{s} = \kappa_{1}\vec{k}_{1}+\kappa_{2}\vec{k}_{2}+\kappa_{3}\vec{k}_{3},\quad
  \omega_{s} = \kappa_{1}\omega_{1}+\kappa_{2}\omega_{2}+\kappa_{3}\omega_{3},
\end{align}
where $\vec{k}_{i}$, $\omega_{i}$ are wavevectors and frequencies of incident pulses and $\kappa_{i} = \pm 1$.
Following common conventions~\cite{Scheurer2001,Khalil2003}, we label the directions associated with $\vec{\kappa}=(\kappa_{1}, \kappa_{2}, \kappa_{3})=(-1, 1, 1)$ as $S_{I}$ (rephasing) and with $(1, -1, 1)$ as $S_{II}$ (non-rephasing).

\begin{figure}[ht]
  \centering
  \includegraphics{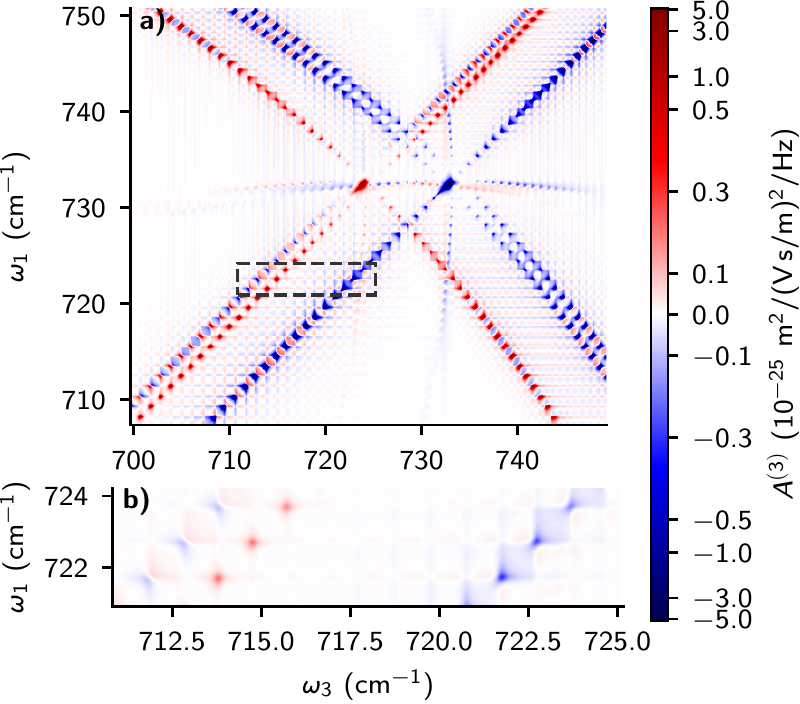}
  \caption{Absorptive (\textit{i.e.} $S_{I} + S_{II}$) 2D spectrum of CH$_{3}$$^{35}$Cl $\nu_3$ mode for waiting time of 1 ps, $T=296$ K and pressure of 1 atmosphere with all polarizations aligned.
  To make narrow resonances more legible, logarithmic scale is used for $|S^{(3)}_{\omega_{3},\omega_{1}}|>0.6$ and linear scale for lower absolute values.
\textbf{b)} Zoom in on the region of the spectrum marked by a dashed rectangle in a).\label{fig:rcsimba}}
\end{figure}

In the impulsive limit (Dirac delta pulses) the third-order signal is given by:
\begin{multline}
  \label{eq:3}
  \hat{\epsilon}_{4}\cdot\vec{P}^{(3)}(z,t_{3}; t_{2}, t_{1}) =\\ i\frac{n\epsilon_{0}c}{\pi\omega_{3}} N \mathcal{E}^{(t)}_{1} \mathcal{E}^{(t)}_{2} \mathcal{E}^{(t)}_{3}
  \sum_{\mathrm{pathways}} e^{i\vec{k_{s}}\cdot{}\vec{r}} S^{(3)} \mathcal{I}(t_{1}, t_{2}, t_{3}),
\end{multline}
where the sum is over all experimentally relevant third-order excitation pathways.
$N$ is the concentration of the active molecule, $\mathcal{E}^{(t)}_{i}$ are the integrated pulse envelopes, and $S^{(3)}$ is the third-order pathway amplitude~\cite{Kowzan2022}.
$\mathcal{I}(t_{1}, t_{2}, t_{3})$ encapsulates molecular dynamics between excitations.
Under a simplified model of gas-phase 2D line shapes, individual resonances are represented by complex 2D Lorentzian profiles~\cite{Hamm2011a,Kowzan2022} that mix absorptive and dispersive parts of the response.

Purely absorptive 2D spectra are commonly obtained by summing rephasing and nonrephasing spectra~\cite{Khalil2003}.
The sum can be easily performed experimentally by using the pump-probe geometry.
Figure~\ref{fig:rcsimba} shows a simulated absorptive spectrum of methyl chloride $\nu_{3}$ mode for beam and detection polarizations all aligned, produced by summing all rovibrational pathways starting in the ground vibrational state with $J$ values up to 30 and calculating 2D Lorentzian profiles for each pathway.
The simulation was performed with our \textsc{rotsim2d} code~\cite{Kowzan2022rotsim2d} using molecular data from the HITRAN database~\cite{Gordon2017,Nikitin2005}.
The plotted quantity is the third-order amplitude
\begin{equation}
  \label{eq:4}
  A^{(3)}(\omega_{1}, t_{2}, \omega_{3}) =  \sum_{\mathrm{pathways}} \mathcal{I}(\omega_{1}, t_{2}, \omega_{3}) S^{(3)}
\end{equation}
at $t_{2}=1$ ps, where $\mathcal{I}(\omega_{1}, t_{2}, \omega_{3})$ is a partial Fourier transform of $\mathcal{I}(t_{1}, t_{2}, t_{3})$. 

In contrast to double-resonance spectroscopy with CW lasers, broadband excitation produces rotational coherences (RC) during the waiting time $t_{2}$.
Inclusion of RC imbalances rephasing and nonrephasing pathways---\textit{i.e.} there are more rephasing pathways than non-rephasing pathways (or \textit{vice versa})---and this prevents acquisition of purely absorptive spectra.
In Fig.~\ref{fig:rcsimba}a), this effect is seen as a faint, diffuse background signal caused by long tails of dispersive line shapes.
Moreover, the positive branches are partially negative and \textit{vice versa}, depending on the degree of imbalance for the specific branch and on the phases accumulated via waiting time, $\Omega_2 t_{2}$, of the contributing RC pathways, with $\Omega_2$ determined by the rotational energy difference between the bra and ket of the coherences \cite{Kowzan2022}. 
The effect on individual resonances is shown in Fig.~\ref{fig:rcsimba}b).
The leftmost resonances belong to a nonrephasing RC branch with no rephasing counterpart and they have purely dispersive shapes.
One progression to the right, there are compact purely absorptive line shapes for which rephasing and nonrephasing non-RC pathways are perfectly balanced.
Finally, the resonances on the main diagonal (rightmost progression) comprise multiple RC and non-RC pathways phasematched in both $S_{I}$ and $S_{II}$ direction.
This produces asymmetric line shapes with the exact shape dependent on $\Omega_2 t_2$, which changes along the branch (see Fig.~\ref{fig:rcsimba}a).

\begin{figure*}[ht]
  \centering	
  \includegraphics{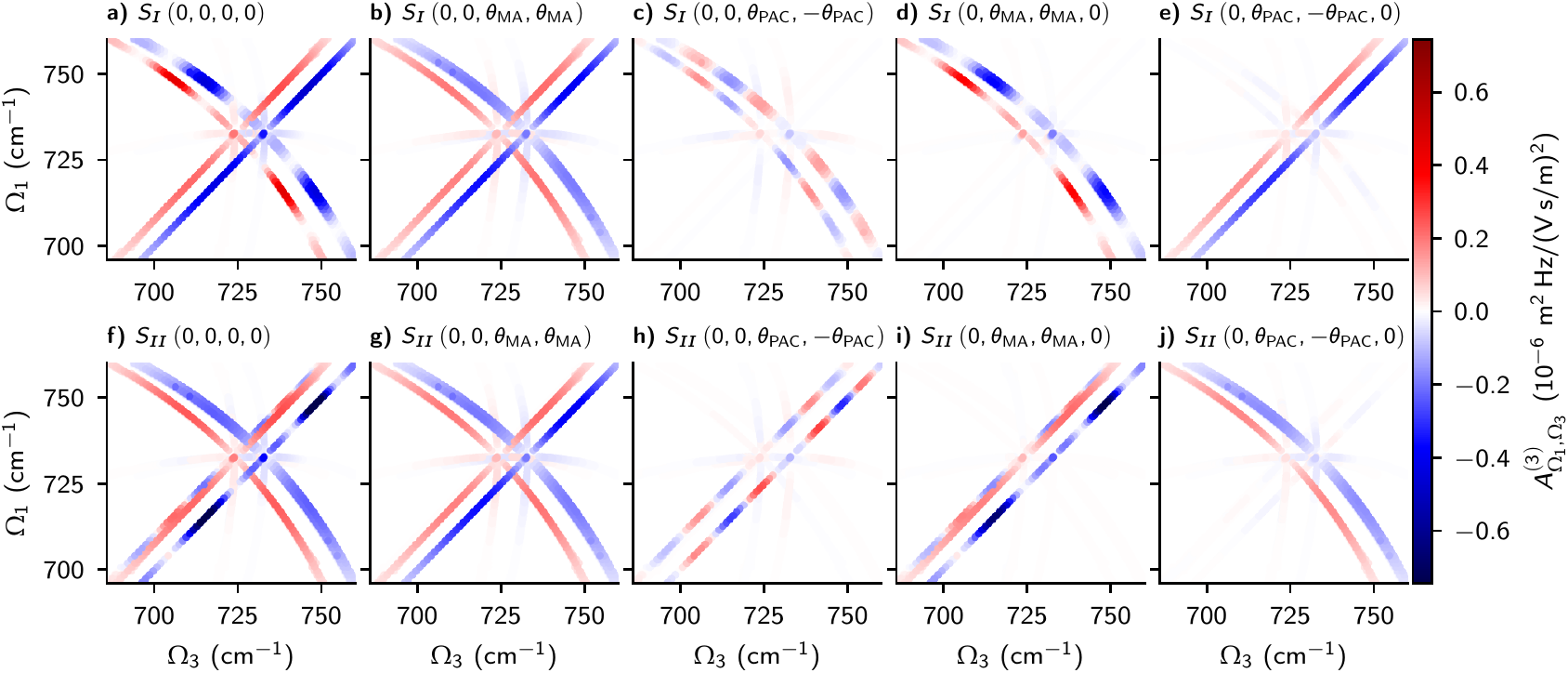}
  \caption{Effect of different polarization conditions on 2D resonance amplitudes, $A^{(3)}_{\Omega_1,\Omega_3}(t_2=1\text{ ps})$.
    Modulation along the branches is caused by rotationally coherent pathways.
  The leftmost column, a) and f), presents the pathway amplitudes for beam and detection polarizations all aligned.
The remaining panels show the effects of four special polarization conditions on $S_{I}$ and $S_{II}$ pathways.}
\label{fig:supgal}
\end{figure*}

Significantly more clarity can be brought to RR2DIR spectra by varying the polarization angles of the three incident pulses $\theta_1, \theta_2$, and $\theta_3$ and also the detected polarization $\theta_4$. 
The polarization-dependence of each pathway in Eq.~\eqref{eq:3} can be derived using spherical tensor operator techniques~\cite{Zare1991,Blum2012}, as demonstrated by Williams~\textit{et al.}~\cite{Williams1994,Williams1994a,Williams1995,Williams1996,Williams1997} and Vaccaro~\textit{et al.}~\cite{Wasserman1998,Bracamonte2003,Murdock2009}.
The key result is the separation of all rovibrational third-order pathways into 7 classes with regards to their dependence on the polarization angles~\cite{Kowzan2022}.
The amplitude of each pathway is proportional to an expression depending only on the beam polarizations, $\bm{\epsilon}=(\theta_{1}, \theta_{2}, \theta_{3}, \theta_{4})$, and on the rotational angular momentum of involved states, $\bm{J}=(J_{i}, J_{j}, J_{k}, J_{l})$, given by~\cite{Murdock2009,Kowzan2022}:
\begin{equation}
  \label{eq:5}
  \begin{split}
    R^{(0)}_{0}(\bm{\epsilon}; \bm{J}) =& \frac{c_{00}}{60(2J_{i}+1)^{3/2}}\big(
    c_{12} \cos(\theta_{1}+\theta_{2}-\theta_{3}-\theta_{4})\\
    &+c_{13} \cos(\theta_{1}-\theta_{2}+\theta_{3}-\theta_{4})\\
    &+c_{14} \cos(\theta_{1}-\theta_{2}-\theta_{3}+\theta_{4})
    \big).
  \end{split}
\end{equation}
Where the $c_{\alpha\beta}$ coefficients in general depend on $\bm{J}$, but in the limit of high $J_{i}$---valid for $J_{i} \gtrsim 10$---they only depend on differences between $J_{\alpha}$ values~\cite{Murdock2009,Kowzan2022}.
This permits derivation of polarization conditions capable of suppressing whole branches of third-order pathways and greatly simplifying RR2DIR spectra. It is easier to illustrate with the $t_2$-dependent 2D resonance amplitudes
\begin{equation}\label{eq:10}
	A^{(3)}_{\Omega_1,\Omega_3}(t_2) = \sum_{\mathrm{pathways}} S^{(3)} e^{-i\Omega_2 t_2} 
\end{equation}
instead of the full spectrum of Fig.~\ref{fig:rcsimba}. Figure~\ref{fig:supgal} shows $A^{(3)}_{\Omega_{1},\Omega_{3}}$ under various polarization conditions, which we explain in the paragraphs below.

A well-known polarization condition in nonlinear spectroscopy is the so-called ``magic angle" (MA) condition where pump and probe pulses have an angle of $\theta_{\mathrm{MA}} = \tan^{-1}\sqrt{2}$ between them, \textit{i.e.} $\bm{\epsilon} = (0,0,\theta_{\mathrm{MA}},\theta_{\mathrm{MA}})$.
The effect of the magic angle condition is often treated classically \cite{Weiner2009, Hamm2011a}.
In the context of 2DIR spectra with resolution of rotational eigenstates, the standard MA condition is more easily understood as a condition that eliminates RC Feynman pathways, as shown in panels b) and g) of Fig.~\ref{fig:supgal}.
The standard MA condition offers some control of RR2DIR spectra and can help with spectral congestion, but all main branches remain intact.

\begin{figure}[h]
  \centering
  \includegraphics{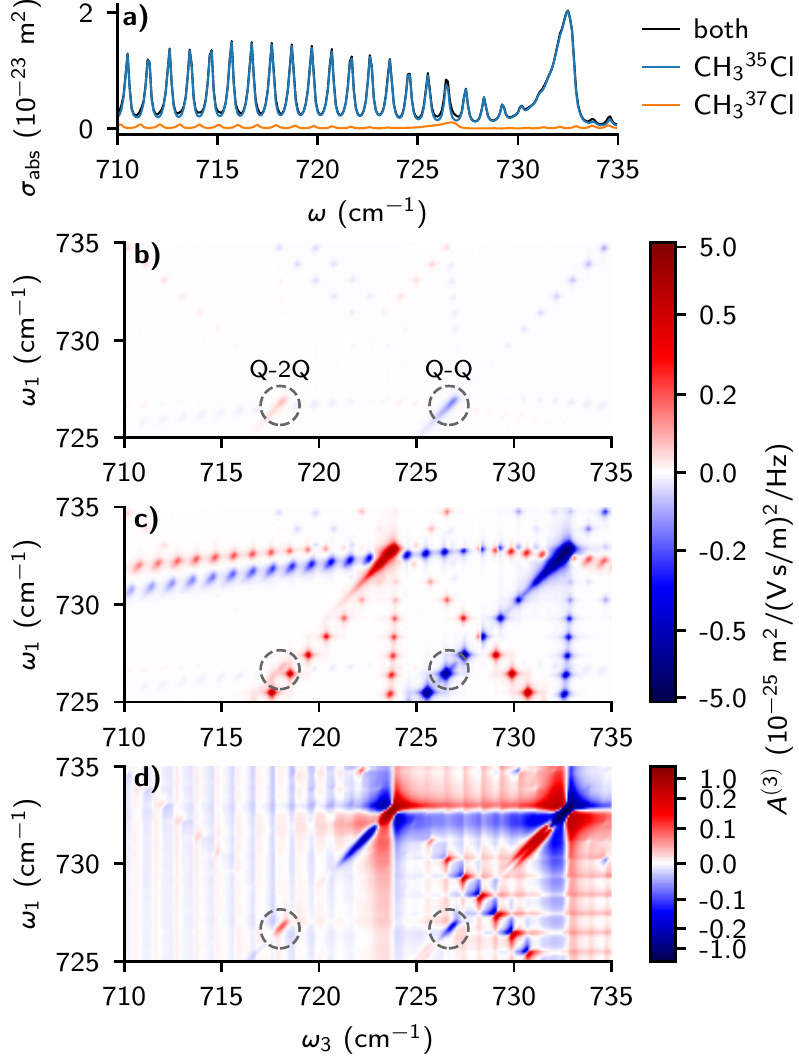}
  \caption{Spectra of 95\% of CH$_{3}$$^{35}$Cl and 5\% of CH$_{3}$$^{37}$Cl at $T=296$ K and pressure of 1 atm.
    Dashed circles mark Q-2Q and Q-Q branches.
\textbf{a)} Linear absorption cross section.
\textbf{b)} 2D absorptive spectrum of 5\% CH$_{3}$$^{37}$Cl under the MA condition at $t_2=1$ ps.
\textbf{c)} 2D absorptive spectrum of both isotopologues under the MA condition at $t_2=1$ ps.
For b) and c) logarithmic scale is used for $|A^{(3)}|>0.58$ and linear scale for lower absolute values.
\textbf{d)} 2D rephasing spectrum of both isotopologues under the middle MA condition at $t_2=1$ ps.
Logarithmic scale is used for $|A^{(3)}|>0.27$ and linear scale for lower absolute values.
\label{fig:symisos}}
\end{figure}

\begin{figure}[ht]
  \centering
  \includegraphics{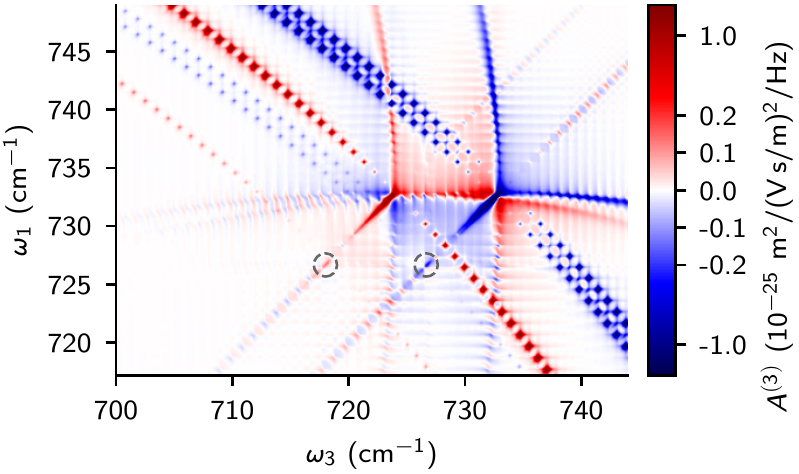}
  \caption{Sum of rephasing spectrum under the middle MA condition and nonrephasing spectrum under the middle PAC condition scaled by $7/3$.
    Dashed circles mark Q-2Q and Q-Q branches.
    Both component spectra were simulated for 95\% of CH$_{3}$$^{35}$Cl and 5\% of CH$_{3}$$^{37}$Cl at $T=296$ K, pressure of 1 atm and $t_2=0$ ps.
    Logarithmic scale is used for $|A^{(3)}|>0.34$ and linear scale for lower absolute values.\label{fig:mapac}}
\end{figure}

For more control of the spectrum, we introduce several new polarization conditions whose effects are illustrated in panels c)--e), h)--j) of Fig.~\ref{fig:supgal}.
For these we introduce the population-alignment canceling (PAC) angle
\begin{equation}
  \label{eq:6}
  \theta_{\mathrm{PAC}}=\sin^{-1} \frac{2}{\sqrt{7}}\approx 49.11^{\circ}\; .
\end{equation}
The arrangement $(0,0,\theta_{\mathrm{PAC}},-\theta_{\mathrm{PAC}})$, illustrated in Fig.~\ref{fig:supgal}c), h) suppresses the non-RC antidiagonal and diagonal branches involving P- and R-type coherences, but leaves unaffected branches involving Q-type coherences.
As a result, this condition brings out the RC pathways in the spectrum and gives a strongly $t_2$-dependent signal.
Another condition we introduce is the middle MA condition, $(0, \theta_{\mathrm{MA}}, \theta_{\mathrm{MA}},0)$, which has some similarities to $(0,0,\theta_{\mathrm{PAC}},-\theta_{\mathrm{PAC}})$, but differs in details.
For $S_{I}$ [Fig.~\ref{fig:supgal}d)], the middle MA condition suppresses off-diagonal branches with Q-type coherences and the diagonal branches, leaving only the anti-diagonal ones and the central Q-Q and Q-2Q branches.
For $S_{II}$ [Fig.~\ref{fig:supgal}i)], the effect of this condition is to remove all non-diagonal branches.
A third new arrangement, the middle PAC condition [Fig.~\ref{fig:supgal}e), j)] attempts to do the opposite of the middle MA condition.
It partially suppresses the antidiagonal resonances for $S_{I}$ and partially the diagonal branches for $S_{II}$. 
As we show in the examples below, using these new polarization conditions can offer dramatic benefits when dealing with RR2DIR spectra of multiple species.

In Figure~\ref{fig:symisos}a) we plot the linear absorption signal for a mixture of 95\% CH$_{3}$$^{35}$Cl and 5\% CH$_{3}$$^{37}$Cl.
The contributions from both isotopologues strongly overlap.
Retrieving component concentrations from such congested linear spectra is a difficult problem in molecular spectroscopy that often requires multispectrum fits~\cite{Benner1995}, sophisticated line shape models~\cite{Hartmann2008}, and careful treatment of systematic effects in the spectrometer, such as etalons, drifting spectral envelope, etc.
In contrast, the two isotopologues can be better separated in RR2DIR spectra.
Figures~\ref{fig:symisos}b) and c) show 2D absorptive spectra under the standard MA condition.
Panel b) shows the weak CH$_{3}$$^{37}$Cl signal with relatively strong Q-2Q and Q-Q branches.
In the combined spectrum, Fig.~\ref{fig:symisos}c), the resonances on the diagonal and shifted diagonal of both isotopologues overlap and are difficult to separate, but the antidiagonal resonances are clearly separated.
The situation improves even more using the middle MA condition, which removes the diagonal branches and isolates the Q-Q and Q-2Q branches of CH$_{3}$$^{37}$Cl from the signals due to the more abundant CH$_{3}$$^{35}$Cl.

Since absorptive 2DIR spectra have the narrowest lineshapes, beter signal separation would be obtained by recording purely absorptive 2D spectra on top of suppressing the diagonal branches, but devising a polarization condition that suppresses the diagonal simulltaneously for both $S_I$ and $S_{II}$ pathways is unfortunately not possible.
This is because the diagonal pathways phase-matched in the $S_{I}$ direction have precisely the same polarization dependence as the antidiagonal pathways emitted in the $S_{II}$ direction~\cite{Kowzan2022} as shown in Fig.~\ref{fig:supgal}. 
Nevertheless, a similar simplification is achieved by summing middle MA rephasing spectrum and middle PAC nonrephasing spectrum scaled by $7/3$, as shown in Fig.~\ref{fig:mapac}.

While other polarization conditions do not improve signal separation so dramatically, their effect can be predicted precisely using Eq.~\eqref{eq:5} and provide constraints on models for fitting multi-species spectra.
Rejecting unphysical models that optimally fit the data is critically important in multi-parameter, multi-component fits~\cite{Benner1995,Stokkum2004,Slavov2015}.
Since the polarization dependence does not distinguish between different internal rotational states, our results are also directly applicable to asymmetric top molecules~\cite{Chew2013} whose eigenfunctions are linear combinations of symmetric top wavefunctions.

Lastly, we estimate experimental detection limits for a standard pump-probe geometry measurement~\cite{Helbing2010} with a Fourier-transform spectrometer.
We assume Fourier transform-limited Gaussian pulses with duration of 300 fs and bandwidth of 50 cm$^{-1}$, producing frequency combs with repetition rate of 100 MHz.
We set the average power of pump beams at 2 W and assume beam radii of 250 $\mu$m.
At $t_{1}=t_{2}=0$, 1 atm. of pure methyl chloride, $T=296$ K, and a path length of 1 cm, the peak pump-induced change in the probe absorbance $1.7\times{}10^{-2}$.
Assuming average probe power of 1 mW and resolution of 1 GHz, the shot-noise-limited, single-element noise-equivalent concentration is $149\text{ ppm}/\sqrt{\text{Hz}}$.
As demonstrated for transient absorption spectroscopy~\cite{Reber2016, Silfies2021}, the detection limits can be improved by $>10^4$ to reach the $5\text{ ppb}/\sqrt{\mbox{Hz}}$ level by using the cavity enhancement schemes proposed in~\cite{Allison2017}.

In summary, RR2DIR spectroscopy has the potential to become a valuable investigative tool in fundamental science and applications.
Adding a second dimension to high resolution spectroscopy offers better discrimination of resonances, which is further enhanced by the presented polarization conditions.
It is also worth noting that since whole branches are affected, high resolution is not required to benefit from polarization control.
In fact, the offered isolation of diagonal, antidiagonal and RC resonances could be particularly illuminating in high-pressure studies of line mixing~\cite{Chen2009} and the gas-to-liquid transition~\cite{Mandal2018,Pack2019}.
With currently available high-power laser technology and optical enhancement cavities, it is feasible to obtain detection limits suitable for trace-gas detection.
This includes investigation of vibrational dynamics in cold molecular beams~\cite{Silfies2021}, as well as applications in breath analysis~\cite{Henderson2018a,Liang2021a}, combustion science~\cite{Chen2007} and plasma diagnostics~\cite{Patnaik2017}.

\begin{acknowledgments}
This project has received funding from the European Union's Horizon 2020 research and innovation programme under the Marie Sklodowska-Curie grant agreement No 101028278. This work was supported by the U.S. National Science Foundation under award number 1708743 and the U.S. Air Force Office of Scientific Research under grant number FA9550-20-1-0259.
\end{acknowledgments}


\begin{thebibliography}{52}%
\makeatletter
\providecommand \@ifxundefined [1]{%
 \@ifx{#1\undefined}
}%
\providecommand \@ifnum [1]{%
 \ifnum #1\expandafter \@firstoftwo
 \else \expandafter \@secondoftwo
 \fi
}%
\providecommand \@ifx [1]{%
 \ifx #1\expandafter \@firstoftwo
 \else \expandafter \@secondoftwo
 \fi
}%
\providecommand \natexlab [1]{#1}%
\providecommand \enquote  [1]{``#1''}%
\providecommand \bibnamefont  [1]{#1}%
\providecommand \bibfnamefont [1]{#1}%
\providecommand \citenamefont [1]{#1}%
\providecommand \href@noop [0]{\@secondoftwo}%
\providecommand \href [0]{\begingroup \@sanitize@url \@href}%
\providecommand \@href[1]{\@@startlink{#1}\@@href}%
\providecommand \@@href[1]{\endgroup#1\@@endlink}%
\providecommand \@sanitize@url [0]{\catcode `\\12\catcode `\$12\catcode
  `\&12\catcode `\#12\catcode `\^12\catcode `\_12\catcode `\%12\relax}%
\providecommand \@@startlink[1]{}%
\providecommand \@@endlink[0]{}%
\providecommand \url  [0]{\begingroup\@sanitize@url \@url }%
\providecommand \@url [1]{\endgroup\@href {#1}{\urlprefix }}%
\providecommand \urlprefix  [0]{URL }%
\providecommand \Eprint [0]{\href }%
\providecommand \doibase [0]{http://dx.doi.org/}%
\providecommand \selectlanguage [0]{\@gobble}%
\providecommand \bibinfo  [0]{\@secondoftwo}%
\providecommand \bibfield  [0]{\@secondoftwo}%
\providecommand \translation [1]{[#1]}%
\providecommand \BibitemOpen [0]{}%
\providecommand \bibitemStop [0]{}%
\providecommand \bibitemNoStop [0]{.\EOS\space}%
\providecommand \EOS [0]{\spacefactor3000\relax}%
\providecommand \BibitemShut  [1]{\csname bibitem#1\endcsname}%
\let\auto@bib@innerbib\@empty
\bibitem [{\citenamefont {Hall}\ \emph {et~al.}(2001)\citenamefont {Hall},
  \citenamefont {Ye}, \citenamefont {Diddams}, \citenamefont {Ma},
  \citenamefont {Cundiff},\ and\ \citenamefont {Jones}}]{Hall2001}%
  \BibitemOpen
  \bibfield  {author} {\bibinfo {author} {\bibfnamefont {J.}~\bibnamefont
  {Hall}}, \bibinfo {author} {\bibfnamefont {J.}~\bibnamefont {Ye}}, \bibinfo
  {author} {\bibfnamefont {S.}~\bibnamefont {Diddams}}, \bibinfo {author}
  {\bibfnamefont {L.-S.}\ \bibnamefont {Ma}}, \bibinfo {author} {\bibfnamefont
  {S.}~\bibnamefont {Cundiff}}, \ and\ \bibinfo {author} {\bibfnamefont
  {D.}~\bibnamefont {Jones}},\ }\href {\doibase 10.1109/3.970893} {\bibfield
  {journal} {\bibinfo  {journal} {IEEE Journal of Quantum Electronics}\
  }\textbf {\bibinfo {volume} {37}},\ \bibinfo {pages} {1482} (\bibinfo {year}
  {2001})}\BibitemShut {NoStop}%
\bibitem [{\citenamefont {Ye}\ and\ \citenamefont {Cundiff}(2005)}]{Ye2005}%
  \BibitemOpen
  \bibinfo {editor} {\bibfnamefont {J.}~\bibnamefont {Ye}}\ and\ \bibinfo
  {editor} {\bibfnamefont {S.~T.}\ \bibnamefont {Cundiff}},\ eds.,\ \href@noop
  {} {{\selectlanguage {english}\emph {\bibinfo {title} {Femtosecond {{Optical
  Frequency Comb}}: {{Principle}}, {{Operation}}, and {{Applications}}}}}}\
  (\bibinfo  {publisher} {{Kluwer Academic Publishers}},\ \bibinfo {address}
  {Boston},\ \bibinfo {year} {2005})\BibitemShut {NoStop}%
\bibitem [{\citenamefont {Kennedy}\ \emph {et~al.}(2020)\citenamefont
  {Kennedy}, \citenamefont {Oelker}, \citenamefont {Robinson}, \citenamefont
  {Bothwell}, \citenamefont {Kedar}, \citenamefont {Milner}, \citenamefont
  {Marti}, \citenamefont {Derevianko},\ and\ \citenamefont {Ye}}]{Kennedy2020}%
  \BibitemOpen
  \bibfield  {author} {\bibinfo {author} {\bibfnamefont {C.~J.}\ \bibnamefont
  {Kennedy}}, \bibinfo {author} {\bibfnamefont {E.}~\bibnamefont {Oelker}},
  \bibinfo {author} {\bibfnamefont {J.~M.}\ \bibnamefont {Robinson}}, \bibinfo
  {author} {\bibfnamefont {T.}~\bibnamefont {Bothwell}}, \bibinfo {author}
  {\bibfnamefont {D.}~\bibnamefont {Kedar}}, \bibinfo {author} {\bibfnamefont
  {W.~R.}\ \bibnamefont {Milner}}, \bibinfo {author} {\bibfnamefont {G.~E.}\
  \bibnamefont {Marti}}, \bibinfo {author} {\bibfnamefont {A.}~\bibnamefont
  {Derevianko}}, \ and\ \bibinfo {author} {\bibfnamefont {J.}~\bibnamefont
  {Ye}},\ }\href {\doibase 10.1103/physrevlett.125.201302} {\bibfield
  {journal} {\bibinfo  {journal} {Phys. Rev. Lett.}\ }\textbf {\bibinfo
  {volume} {125}} (\bibinfo {year} {2020}),\
  10.1103/physrevlett.125.201302}\BibitemShut {NoStop}%
\bibitem [{\citenamefont {Steinmetz}\ \emph {et~al.}(2008)\citenamefont
  {Steinmetz}, \citenamefont {Wilken}, \citenamefont {{Araujo-Hauck}},
  \citenamefont {Holzwarth}, \citenamefont {H{\"a}nsch}, \citenamefont
  {Pasquini}, \citenamefont {Manescau}, \citenamefont {D'Odorico},
  \citenamefont {Murphy}, \citenamefont {Kentischer}, \citenamefont {Schmidt},\
  and\ \citenamefont {Udem}}]{Steinmetz2008}%
  \BibitemOpen
  \bibfield  {author} {\bibinfo {author} {\bibfnamefont {T.}~\bibnamefont
  {Steinmetz}}, \bibinfo {author} {\bibfnamefont {T.}~\bibnamefont {Wilken}},
  \bibinfo {author} {\bibfnamefont {C.}~\bibnamefont {{Araujo-Hauck}}},
  \bibinfo {author} {\bibfnamefont {R.}~\bibnamefont {Holzwarth}}, \bibinfo
  {author} {\bibfnamefont {T.~W.}\ \bibnamefont {H{\"a}nsch}}, \bibinfo
  {author} {\bibfnamefont {L.}~\bibnamefont {Pasquini}}, \bibinfo {author}
  {\bibfnamefont {A.}~\bibnamefont {Manescau}}, \bibinfo {author}
  {\bibfnamefont {S.}~\bibnamefont {D'Odorico}}, \bibinfo {author}
  {\bibfnamefont {M.~T.}\ \bibnamefont {Murphy}}, \bibinfo {author}
  {\bibfnamefont {T.}~\bibnamefont {Kentischer}}, \bibinfo {author}
  {\bibfnamefont {W.}~\bibnamefont {Schmidt}}, \ and\ \bibinfo {author}
  {\bibfnamefont {T.}~\bibnamefont {Udem}},\ }\href {\doibase
  10.1126/science.1161030} {\bibfield  {journal} {\bibinfo  {journal}
  {Science}\ }\textbf {\bibinfo {volume} {321}},\ \bibinfo {pages} {1335}
  (\bibinfo {year} {2008})}\BibitemShut {NoStop}%
\bibitem [{\citenamefont {Suh}\ \emph {et~al.}(2018)\citenamefont {Suh},
  \citenamefont {Yi}, \citenamefont {Lai}, \citenamefont {Leifer},
  \citenamefont {Grudinin}, \citenamefont {Vasisht}, \citenamefont {Martin},
  \citenamefont {Fitzgerald}, \citenamefont {Doppmann}, \citenamefont {Wang},
  \citenamefont {Mawet}, \citenamefont {Papp}, \citenamefont {Diddams},
  \citenamefont {Beichman},\ and\ \citenamefont {Vahala}}]{Suh2018a}%
  \BibitemOpen
  \bibfield  {author} {\bibinfo {author} {\bibfnamefont {M.-G.}\ \bibnamefont
  {Suh}}, \bibinfo {author} {\bibfnamefont {X.}~\bibnamefont {Yi}}, \bibinfo
  {author} {\bibfnamefont {Y.-H.}\ \bibnamefont {Lai}}, \bibinfo {author}
  {\bibfnamefont {S.}~\bibnamefont {Leifer}}, \bibinfo {author} {\bibfnamefont
  {I.~S.}\ \bibnamefont {Grudinin}}, \bibinfo {author} {\bibfnamefont
  {G.}~\bibnamefont {Vasisht}}, \bibinfo {author} {\bibfnamefont {E.~C.}\
  \bibnamefont {Martin}}, \bibinfo {author} {\bibfnamefont {M.~P.}\
  \bibnamefont {Fitzgerald}}, \bibinfo {author} {\bibfnamefont
  {G.}~\bibnamefont {Doppmann}}, \bibinfo {author} {\bibfnamefont
  {J.}~\bibnamefont {Wang}}, \bibinfo {author} {\bibfnamefont {D.}~\bibnamefont
  {Mawet}}, \bibinfo {author} {\bibfnamefont {S.~B.}\ \bibnamefont {Papp}},
  \bibinfo {author} {\bibfnamefont {S.~A.}\ \bibnamefont {Diddams}}, \bibinfo
  {author} {\bibfnamefont {C.}~\bibnamefont {Beichman}}, \ and\ \bibinfo
  {author} {\bibfnamefont {K.}~\bibnamefont {Vahala}},\ }\href {\doibase
  10.1038/s41566-018-0312-3} {\bibfield  {journal} {\bibinfo  {journal} {Nat.
  Photonics}\ }\textbf {\bibinfo {volume} {13}},\ \bibinfo {pages} {25–30}
  (\bibinfo {year} {2018})}\BibitemShut {NoStop}%
\bibitem [{\citenamefont {Weichman}\ \emph {et~al.}(2019)\citenamefont
  {Weichman}, \citenamefont {Changala}, \citenamefont {Ye}, \citenamefont
  {Chen}, \citenamefont {Yan},\ and\ \citenamefont
  {Picqu{\'e}}}]{Weichman2019}%
  \BibitemOpen
  \bibfield  {author} {\bibinfo {author} {\bibfnamefont {M.~L.}\ \bibnamefont
  {Weichman}}, \bibinfo {author} {\bibfnamefont {P.~B.}\ \bibnamefont
  {Changala}}, \bibinfo {author} {\bibfnamefont {J.}~\bibnamefont {Ye}},
  \bibinfo {author} {\bibfnamefont {Z.}~\bibnamefont {Chen}}, \bibinfo {author}
  {\bibfnamefont {M.}~\bibnamefont {Yan}}, \ and\ \bibinfo {author}
  {\bibfnamefont {N.}~\bibnamefont {Picqu{\'e}}},\ }\href {\doibase
  10.1016/j.jms.2018.11.011} {\bibfield  {journal} {\bibinfo  {journal} {J.
  Mol. Spectrosc.}\ }\textbf {\bibinfo {volume} {355}},\ \bibinfo {pages} {66}
  (\bibinfo {year} {2019})}\BibitemShut {NoStop}%
\bibitem [{\citenamefont {Coddington}\ \emph {et~al.}(2016)\citenamefont
  {Coddington}, \citenamefont {Newbury},\ and\ \citenamefont
  {Swann}}]{Coddington2016}%
  \BibitemOpen
  \bibfield  {author} {\bibinfo {author} {\bibfnamefont {I.}~\bibnamefont
  {Coddington}}, \bibinfo {author} {\bibfnamefont {N.}~\bibnamefont {Newbury}},
  \ and\ \bibinfo {author} {\bibfnamefont {W.}~\bibnamefont {Swann}},\ }\href
  {\doibase 10.1364/OPTICA.3.000414} {\bibfield  {journal} {\bibinfo  {journal}
  {Optica}\ }\textbf {\bibinfo {volume} {3}},\ \bibinfo {pages} {414} (\bibinfo
  {year} {2016})}\BibitemShut {NoStop}%
\bibitem [{\citenamefont {Cing{\"o}z}\ \emph {et~al.}(2012)\citenamefont
  {Cing{\"o}z}, \citenamefont {Yost}, \citenamefont {Allison}, \citenamefont
  {Ruehl}, \citenamefont {Fermann}, \citenamefont {Hartl},\ and\ \citenamefont
  {Ye}}]{Cingoz2012}%
  \BibitemOpen
  \bibfield  {author} {\bibinfo {author} {\bibfnamefont {A.}~\bibnamefont
  {Cing{\"o}z}}, \bibinfo {author} {\bibfnamefont {D.~C.}\ \bibnamefont
  {Yost}}, \bibinfo {author} {\bibfnamefont {T.~K.}\ \bibnamefont {Allison}},
  \bibinfo {author} {\bibfnamefont {A.}~\bibnamefont {Ruehl}}, \bibinfo
  {author} {\bibfnamefont {M.~E.}\ \bibnamefont {Fermann}}, \bibinfo {author}
  {\bibfnamefont {I.}~\bibnamefont {Hartl}}, \ and\ \bibinfo {author}
  {\bibfnamefont {J.}~\bibnamefont {Ye}},\ }\href@noop {} {\bibfield  {journal}
  {\bibinfo  {journal} {Nature}\ }\textbf {\bibinfo {volume} {482}},\ \bibinfo
  {pages} {68} (\bibinfo {year} {2012})}\BibitemShut {NoStop}%
\bibitem [{\citenamefont {Liang}\ \emph {et~al.}(2022)\citenamefont {Liang},
  \citenamefont {Chan}, \citenamefont {Toscano}, \citenamefont {Bjorkman},
  \citenamefont {Leinwand}, \citenamefont {Parker}, \citenamefont {Nesbitt},\
  and\ \citenamefont {Ye}}]{Liang2022}%
  \BibitemOpen
  \bibfield  {author} {\bibinfo {author} {\bibfnamefont {Q.}~\bibnamefont
  {Liang}}, \bibinfo {author} {\bibfnamefont {Y.-C.}\ \bibnamefont {Chan}},
  \bibinfo {author} {\bibfnamefont {J.}~\bibnamefont {Toscano}}, \bibinfo
  {author} {\bibfnamefont {K.~K.}\ \bibnamefont {Bjorkman}}, \bibinfo {author}
  {\bibfnamefont {L.~A.}\ \bibnamefont {Leinwand}}, \bibinfo {author}
  {\bibfnamefont {R.}~\bibnamefont {Parker}}, \bibinfo {author} {\bibfnamefont
  {D.~J.}\ \bibnamefont {Nesbitt}}, \ and\ \bibinfo {author} {\bibfnamefont
  {J.}~\bibnamefont {Ye}},\ }\href {http://arxiv.org/abs/2202.02321} {\bibfield
   {journal} {\bibinfo  {journal} {arXiv:2202.02321 [physics]}\ } (\bibinfo
  {year} {2022})},\ \bibinfo {note} {arXiv: 2202.02321}\BibitemShut {NoStop}%
\bibitem [{\citenamefont {Reber}\ \emph {et~al.}(2016)\citenamefont {Reber},
  \citenamefont {Chen},\ and\ \citenamefont {Allison}}]{Reber2016}%
  \BibitemOpen
  \bibfield  {author} {\bibinfo {author} {\bibfnamefont {M.~A.~R.}\
  \bibnamefont {Reber}}, \bibinfo {author} {\bibfnamefont {Y.}~\bibnamefont
  {Chen}}, \ and\ \bibinfo {author} {\bibfnamefont {T.~K.}\ \bibnamefont
  {Allison}},\ }\href {\doibase 10.1364/OPTICA.3.000311} {\bibfield  {journal}
  {\bibinfo  {journal} {Optica}\ }\textbf {\bibinfo {volume} {3}},\ \bibinfo
  {pages} {311} (\bibinfo {year} {2016})}\BibitemShut {NoStop}%
\bibitem [{\citenamefont {Silfies}\ \emph {et~al.}(2021)\citenamefont
  {Silfies}, \citenamefont {Kowzan}, \citenamefont {Lewis},\ and\ \citenamefont
  {Allison}}]{Silfies2021}%
  \BibitemOpen
  \bibfield  {author} {\bibinfo {author} {\bibfnamefont {M.~C.}\ \bibnamefont
  {Silfies}}, \bibinfo {author} {\bibfnamefont {G.}~\bibnamefont {Kowzan}},
  \bibinfo {author} {\bibfnamefont {N.}~\bibnamefont {Lewis}}, \ and\ \bibinfo
  {author} {\bibfnamefont {T.~K.}\ \bibnamefont {Allison}},\ }\href {\doibase
  10.1039/D1CP00631B} {\bibfield  {journal} {\bibinfo  {journal} {Phys. Chem.
  Chem. Phys.}\ }\textbf {\bibinfo {volume} {23}},\ \bibinfo {pages} {9743}
  (\bibinfo {year} {2021})}\BibitemShut {NoStop}%
\bibitem [{\citenamefont {Allison}(2017)}]{Allison2017}%
  \BibitemOpen
  \bibfield  {author} {\bibinfo {author} {\bibfnamefont {T.~K.}\ \bibnamefont
  {Allison}},\ }\href {\doibase 10.1088/1361-6455/50/4/044004} {\bibfield
  {journal} {\bibinfo  {journal} {J. Phys. B: At., Mol. Opt. Phys.}\ }\textbf
  {\bibinfo {volume} {50}},\ \bibinfo {pages} {044004} (\bibinfo {year}
  {2017})}\BibitemShut {NoStop}%
\bibitem [{\citenamefont {Lomsadze}\ and\ \citenamefont
  {Cundiff}(2018)}]{Lomsadze2018a}%
  \BibitemOpen
  \bibfield  {author} {\bibinfo {author} {\bibfnamefont {B.}~\bibnamefont
  {Lomsadze}}\ and\ \bibinfo {author} {\bibfnamefont {S.~T.}\ \bibnamefont
  {Cundiff}},\ }\href {\doibase 10.1103/PhysRevLett.120.233401} {\bibfield
  {journal} {\bibinfo  {journal} {Phys. Rev. Lett.}\ }\textbf {\bibinfo
  {volume} {120}},\ \bibinfo {pages} {233401} (\bibinfo {year}
  {2018})}\BibitemShut {NoStop}%
\bibitem [{\citenamefont {Foltynowicz}\ \emph {et~al.}(2021)\citenamefont
  {Foltynowicz}, \citenamefont {Rutkowski}, \citenamefont {Silander},
  \citenamefont {Johansson}, \citenamefont {de~Oliveira}, \citenamefont
  {Axner}, \citenamefont {Soboń}, \citenamefont {Martynkien}, \citenamefont
  {Mergo},\ and\ \citenamefont {Lehmann}}]{Foltynowicz2021}%
  \BibitemOpen
  \bibfield  {author} {\bibinfo {author} {\bibfnamefont {A.}~\bibnamefont
  {Foltynowicz}}, \bibinfo {author} {\bibfnamefont {L.}~\bibnamefont
  {Rutkowski}}, \bibinfo {author} {\bibfnamefont {I.}~\bibnamefont {Silander}},
  \bibinfo {author} {\bibfnamefont {A.~C.}\ \bibnamefont {Johansson}}, \bibinfo
  {author} {\bibfnamefont {V.~S.}\ \bibnamefont {de~Oliveira}}, \bibinfo
  {author} {\bibfnamefont {O.}~\bibnamefont {Axner}}, \bibinfo {author}
  {\bibfnamefont {G.}~\bibnamefont {Soboń}}, \bibinfo {author} {\bibfnamefont
  {T.}~\bibnamefont {Martynkien}}, \bibinfo {author} {\bibfnamefont
  {P.}~\bibnamefont {Mergo}}, \ and\ \bibinfo {author} {\bibfnamefont {K.~K.}\
  \bibnamefont {Lehmann}},\ }\href {\doibase 10.1103/physrevlett.126.063001}
  {\bibfield  {journal} {\bibinfo  {journal} {Phys. Rev. Lett.}\ }\textbf
  {\bibinfo {volume} {126}},\ \bibinfo {pages} {063001} (\bibinfo {year}
  {2021})}\BibitemShut {NoStop}%
\bibitem [{\citenamefont {Catanese}\ \emph {et~al.}(2020)\citenamefont
  {Catanese}, \citenamefont {Rutledge}, \citenamefont {Silfies}, \citenamefont
  {Li}, \citenamefont {Timmers}, \citenamefont {Kowligy}, \citenamefont {Lind},
  \citenamefont {Diddams},\ and\ \citenamefont {Allison}}]{Catanese2020}%
  \BibitemOpen
  \bibfield  {author} {\bibinfo {author} {\bibfnamefont {A.}~\bibnamefont
  {Catanese}}, \bibinfo {author} {\bibfnamefont {J.}~\bibnamefont {Rutledge}},
  \bibinfo {author} {\bibfnamefont {M.~C.}\ \bibnamefont {Silfies}}, \bibinfo
  {author} {\bibfnamefont {X.}~\bibnamefont {Li}}, \bibinfo {author}
  {\bibfnamefont {H.}~\bibnamefont {Timmers}}, \bibinfo {author} {\bibfnamefont
  {A.~S.}\ \bibnamefont {Kowligy}}, \bibinfo {author} {\bibfnamefont
  {A.}~\bibnamefont {Lind}}, \bibinfo {author} {\bibfnamefont {S.~A.}\
  \bibnamefont {Diddams}}, \ and\ \bibinfo {author} {\bibfnamefont {T.~K.}\
  \bibnamefont {Allison}},\ }\href {\doibase 10.1364/ol.385294} {\bibfield
  {journal} {\bibinfo  {journal} {Opt. Lett.}\ }\textbf {\bibinfo {volume}
  {45}},\ \bibinfo {pages} {1248} (\bibinfo {year} {2020})}\BibitemShut
  {NoStop}%
\bibitem [{\citenamefont {Xing}\ \emph {et~al.}(2021)\citenamefont {Xing},
  \citenamefont {Lesko}, \citenamefont {Umeki}, \citenamefont {Lind},
  \citenamefont {Hoghooghi}, \citenamefont {Wu},\ and\ \citenamefont
  {Diddams}}]{Xing2021}%
  \BibitemOpen
  \bibfield  {author} {\bibinfo {author} {\bibfnamefont {S.}~\bibnamefont
  {Xing}}, \bibinfo {author} {\bibfnamefont {D.~M.~B.}\ \bibnamefont {Lesko}},
  \bibinfo {author} {\bibfnamefont {T.}~\bibnamefont {Umeki}}, \bibinfo
  {author} {\bibfnamefont {A.~J.}\ \bibnamefont {Lind}}, \bibinfo {author}
  {\bibfnamefont {N.}~\bibnamefont {Hoghooghi}}, \bibinfo {author}
  {\bibfnamefont {T.-H.}\ \bibnamefont {Wu}}, \ and\ \bibinfo {author}
  {\bibfnamefont {S.~A.}\ \bibnamefont {Diddams}},\ }\href {\doibase
  10.1063/5.0055534} {\bibfield  {journal} {\bibinfo  {journal} {APL
  Photonics}\ }\textbf {\bibinfo {volume} {6}},\ \bibinfo {pages} {086110}
  (\bibinfo {year} {2021})}\BibitemShut {NoStop}%
\bibitem [{\citenamefont {Nakamura}\ \emph {et~al.}(2022)\citenamefont
  {Nakamura}, \citenamefont {Badarla}, \citenamefont {Hashimoto}, \citenamefont
  {Schunemann},\ and\ \citenamefont {Ideguchi}}]{Nakamura2022}%
  \BibitemOpen
  \bibfield  {author} {\bibinfo {author} {\bibfnamefont {T.}~\bibnamefont
  {Nakamura}}, \bibinfo {author} {\bibfnamefont {V.~R.}\ \bibnamefont
  {Badarla}}, \bibinfo {author} {\bibfnamefont {K.}~\bibnamefont {Hashimoto}},
  \bibinfo {author} {\bibfnamefont {P.~G.}\ \bibnamefont {Schunemann}}, \ and\
  \bibinfo {author} {\bibfnamefont {T.}~\bibnamefont {Ideguchi}},\ }\href
  {\doibase 10.1364/OL.450921} {\bibfield  {journal} {\bibinfo  {journal} {Opt.
  Lett.}\ }\textbf {\bibinfo {volume} {47}},\ \bibinfo {pages} {1790} (\bibinfo
  {year} {2022})}\BibitemShut {NoStop}%
\bibitem [{\citenamefont {Ru}\ \emph {et~al.}(2021)\citenamefont {Ru},
  \citenamefont {Kawamori}, \citenamefont {Schunemann}, \citenamefont
  {Vasilyev}, \citenamefont {Mirov},\ and\ \citenamefont
  {Vodopyanov}}]{Ru2021}%
  \BibitemOpen
  \bibfield  {author} {\bibinfo {author} {\bibfnamefont {Q.}~\bibnamefont
  {Ru}}, \bibinfo {author} {\bibfnamefont {T.}~\bibnamefont {Kawamori}},
  \bibinfo {author} {\bibfnamefont {P.~G.}\ \bibnamefont {Schunemann}},
  \bibinfo {author} {\bibfnamefont {S.}~\bibnamefont {Vasilyev}}, \bibinfo
  {author} {\bibfnamefont {S.~B.}\ \bibnamefont {Mirov}}, \ and\ \bibinfo
  {author} {\bibfnamefont {K.~L.}\ \bibnamefont {Vodopyanov}},\ }\href
  {\doibase 10.1364/ol.403910} {\bibfield  {journal} {\bibinfo  {journal} {Opt.
  Lett.}\ }\textbf {\bibinfo {volume} {46}},\ \bibinfo {pages} {709} (\bibinfo
  {year} {2021})}\BibitemShut {NoStop}%
\bibitem [{\citenamefont {Gaida}\ \emph {et~al.}(2018)\citenamefont {Gaida},
  \citenamefont {Gebhardt}, \citenamefont {Heuermann}, \citenamefont {Stutzki},
  \citenamefont {Jauregui}, \citenamefont {Antonio-Lopez}, \citenamefont
  {Schülzgen}, \citenamefont {Amezcua-Correa}, \citenamefont {Tünnermann},
  \citenamefont {Pupeza},\ and\ \citenamefont {et~al.}}]{Gaida2018b}%
  \BibitemOpen
  \bibfield  {author} {\bibinfo {author} {\bibfnamefont {C.}~\bibnamefont
  {Gaida}}, \bibinfo {author} {\bibfnamefont {M.}~\bibnamefont {Gebhardt}},
  \bibinfo {author} {\bibfnamefont {T.}~\bibnamefont {Heuermann}}, \bibinfo
  {author} {\bibfnamefont {F.}~\bibnamefont {Stutzki}}, \bibinfo {author}
  {\bibfnamefont {C.}~\bibnamefont {Jauregui}}, \bibinfo {author}
  {\bibfnamefont {J.}~\bibnamefont {Antonio-Lopez}}, \bibinfo {author}
  {\bibfnamefont {A.}~\bibnamefont {Schülzgen}}, \bibinfo {author}
  {\bibfnamefont {R.}~\bibnamefont {Amezcua-Correa}}, \bibinfo {author}
  {\bibfnamefont {A.}~\bibnamefont {Tünnermann}}, \bibinfo {author}
  {\bibfnamefont {I.}~\bibnamefont {Pupeza}}, \ and\ \bibinfo {author}
  {\bibnamefont {et~al.}},\ }\href {\doibase 10.1038/s41377-018-0099-5}
  {\bibfield  {journal} {\bibinfo  {journal} {Light: Science \& Applications}\
  }\textbf {\bibinfo {volume} {7}} (\bibinfo {year} {2018}),\
  10.1038/s41377-018-0099-5}\BibitemShut {NoStop}%
\bibitem [{\citenamefont {Lesko}\ \emph {et~al.}(2021)\citenamefont {Lesko},
  \citenamefont {Timmers}, \citenamefont {Xing}, \citenamefont {Kowligy},
  \citenamefont {Lind},\ and\ \citenamefont {Diddams}}]{Lesko2021}%
  \BibitemOpen
  \bibfield  {author} {\bibinfo {author} {\bibfnamefont {D.~M.~B.}\
  \bibnamefont {Lesko}}, \bibinfo {author} {\bibfnamefont {H.}~\bibnamefont
  {Timmers}}, \bibinfo {author} {\bibfnamefont {S.}~\bibnamefont {Xing}},
  \bibinfo {author} {\bibfnamefont {A.}~\bibnamefont {Kowligy}}, \bibinfo
  {author} {\bibfnamefont {A.~J.}\ \bibnamefont {Lind}}, \ and\ \bibinfo
  {author} {\bibfnamefont {S.~A.}\ \bibnamefont {Diddams}},\ }\href {\doibase
  10.1038/s41566-021-00778-y} {\bibfield  {journal} {\bibinfo  {journal} {Nat.
  Photonics}\ }\textbf {\bibinfo {volume} {15}},\ \bibinfo {pages} {281}
  (\bibinfo {year} {2021})}\BibitemShut {NoStop}%
\bibitem [{\citenamefont {Mandal}\ \emph {et~al.}(2018)\citenamefont {Mandal},
  \citenamefont {Pack}, \citenamefont {Shah}, \citenamefont {Erramilli},\ and\
  \citenamefont {Ziegler}}]{Mandal2018}%
  \BibitemOpen
  \bibfield  {author} {\bibinfo {author} {\bibfnamefont {A.}~\bibnamefont
  {Mandal}}, \bibinfo {author} {\bibfnamefont {G.~N.}\ \bibnamefont {Pack}},
  \bibinfo {author} {\bibfnamefont {P.~P.}\ \bibnamefont {Shah}}, \bibinfo
  {author} {\bibfnamefont {S.}~\bibnamefont {Erramilli}}, \ and\ \bibinfo
  {author} {\bibfnamefont {L.~D.}\ \bibnamefont {Ziegler}},\ }\href {\doibase
  10.1103/physrevlett.120.103401} {\bibfield  {journal} {\bibinfo  {journal}
  {Phys. Rev. Lett.}\ }\textbf {\bibinfo {volume} {120}},\ \bibinfo {pages}
  {103401} (\bibinfo {year} {2018})}\BibitemShut {NoStop}%
\bibitem [{\citenamefont {Williams}\ \emph
  {et~al.}(1994{\natexlab{a}})\citenamefont {Williams}, \citenamefont {Zare},\
  and\ \citenamefont {Rahn}}]{Williams1994}%
  \BibitemOpen
  \bibfield  {author} {\bibinfo {author} {\bibfnamefont {S.}~\bibnamefont
  {Williams}}, \bibinfo {author} {\bibfnamefont {R.~N.}\ \bibnamefont {Zare}},
  \ and\ \bibinfo {author} {\bibfnamefont {L.~A.}\ \bibnamefont {Rahn}},\
  }\href {\doibase 10.1063/1.467804} {\bibfield  {journal} {\bibinfo  {journal}
  {J. Chem. Phys.}\ }\textbf {\bibinfo {volume} {101}},\ \bibinfo {pages}
  {1072} (\bibinfo {year} {1994}{\natexlab{a}})}\BibitemShut {NoStop}%
\bibitem [{\citenamefont {Williams}\ \emph
  {et~al.}(1994{\natexlab{b}})\citenamefont {Williams}, \citenamefont {Zare},\
  and\ \citenamefont {Rahn}}]{Williams1994a}%
  \BibitemOpen
  \bibfield  {author} {\bibinfo {author} {\bibfnamefont {S.}~\bibnamefont
  {Williams}}, \bibinfo {author} {\bibfnamefont {R.~N.}\ \bibnamefont {Zare}},
  \ and\ \bibinfo {author} {\bibfnamefont {L.~A.}\ \bibnamefont {Rahn}},\
  }\href {\doibase 10.1063/1.467805} {\bibfield  {journal} {\bibinfo  {journal}
  {J. Chem. Phys.}\ }\textbf {\bibinfo {volume} {101}},\ \bibinfo {pages}
  {1093} (\bibinfo {year} {1994}{\natexlab{b}})}\BibitemShut {NoStop}%
\bibitem [{\citenamefont {Williams}\ \emph {et~al.}(1995)\citenamefont
  {Williams}, \citenamefont {Tobiason}, \citenamefont {Dunlop},\ and\
  \citenamefont {Rohlfing}}]{Williams1995}%
  \BibitemOpen
  \bibfield  {author} {\bibinfo {author} {\bibfnamefont {S.}~\bibnamefont
  {Williams}}, \bibinfo {author} {\bibfnamefont {J.~D.}\ \bibnamefont
  {Tobiason}}, \bibinfo {author} {\bibfnamefont {J.~R.}\ \bibnamefont
  {Dunlop}}, \ and\ \bibinfo {author} {\bibfnamefont {E.~A.}\ \bibnamefont
  {Rohlfing}},\ }\href {\doibase 10.1063/1.468826} {\bibfield  {journal}
  {\bibinfo  {journal} {J. Chem. Phys.}\ }\textbf {\bibinfo {volume} {102}},\
  \bibinfo {pages} {8342} (\bibinfo {year} {1995})}\BibitemShut {NoStop}%
\bibitem [{\citenamefont {Williams}\ \emph {et~al.}(1996)\citenamefont
  {Williams}, \citenamefont {Rahn},\ and\ \citenamefont {Zare}}]{Williams1996}%
  \BibitemOpen
  \bibfield  {author} {\bibinfo {author} {\bibfnamefont {S.}~\bibnamefont
  {Williams}}, \bibinfo {author} {\bibfnamefont {L.~A.}\ \bibnamefont {Rahn}},
  \ and\ \bibinfo {author} {\bibfnamefont {R.~N.}\ \bibnamefont {Zare}},\
  }\href {\doibase 10.1063/1.471250} {\bibfield  {journal} {\bibinfo  {journal}
  {J. Chem. Phys.}\ }\textbf {\bibinfo {volume} {104}},\ \bibinfo {pages}
  {3947} (\bibinfo {year} {1996})}\BibitemShut {NoStop}%
\bibitem [{\citenamefont {Williams}\ \emph {et~al.}(1997)\citenamefont
  {Williams}, \citenamefont {Rohlfing}, \citenamefont {Rahn},\ and\
  \citenamefont {Zare}}]{Williams1997}%
  \BibitemOpen
  \bibfield  {author} {\bibinfo {author} {\bibfnamefont {S.}~\bibnamefont
  {Williams}}, \bibinfo {author} {\bibfnamefont {E.~A.}\ \bibnamefont
  {Rohlfing}}, \bibinfo {author} {\bibfnamefont {L.~A.}\ \bibnamefont {Rahn}},
  \ and\ \bibinfo {author} {\bibfnamefont {R.~N.}\ \bibnamefont {Zare}},\
  }\href {\doibase 10.1063/1.473052} {\bibfield  {journal} {\bibinfo  {journal}
  {J. Chem. Phys.}\ }\textbf {\bibinfo {volume} {106}},\ \bibinfo {pages}
  {3090} (\bibinfo {year} {1997})}\BibitemShut {NoStop}%
\bibitem [{\citenamefont {Wasserman}\ \emph {et~al.}(1998)\citenamefont
  {Wasserman}, \citenamefont {Vaccaro},\ and\ \citenamefont
  {Johnson}}]{Wasserman1998}%
  \BibitemOpen
  \bibfield  {author} {\bibinfo {author} {\bibfnamefont {T.~A.~W.}\
  \bibnamefont {Wasserman}}, \bibinfo {author} {\bibfnamefont {P.~H.}\
  \bibnamefont {Vaccaro}}, \ and\ \bibinfo {author} {\bibfnamefont {B.~R.}\
  \bibnamefont {Johnson}},\ }\href {\doibase 10.1063/1.476207} {\bibfield
  {journal} {\bibinfo  {journal} {J. Chem. Phys.}\ }\textbf {\bibinfo {volume}
  {108}},\ \bibinfo {pages} {7713} (\bibinfo {year} {1998})}\BibitemShut
  {NoStop}%
\bibitem [{\citenamefont {Bracamonte}\ and\ \citenamefont
  {Vaccaro}(2003)}]{Bracamonte2003}%
  \BibitemOpen
  \bibfield  {author} {\bibinfo {author} {\bibfnamefont {A.~E.}\ \bibnamefont
  {Bracamonte}}\ and\ \bibinfo {author} {\bibfnamefont {P.~H.}\ \bibnamefont
  {Vaccaro}},\ }\href {\doibase 10.1063/1.1579472} {\bibfield  {journal}
  {\bibinfo  {journal} {J. Chem. Phys.}\ }\textbf {\bibinfo {volume} {119}},\
  \bibinfo {pages} {887} (\bibinfo {year} {2003})}\BibitemShut {NoStop}%
\bibitem [{\citenamefont {Murdock}\ \emph {et~al.}(2009)\citenamefont
  {Murdock}, \citenamefont {Burns},\ and\ \citenamefont
  {Vaccaro}}]{Murdock2009}%
  \BibitemOpen
  \bibfield  {author} {\bibinfo {author} {\bibfnamefont {D.}~\bibnamefont
  {Murdock}}, \bibinfo {author} {\bibfnamefont {L.~A.}\ \bibnamefont {Burns}},
  \ and\ \bibinfo {author} {\bibfnamefont {P.~H.}\ \bibnamefont {Vaccaro}},\
  }\href {\doibase 10.1021/jp903970d} {\bibfield  {journal} {\bibinfo
  {journal} {J. Phys. Chem. A}\ }\textbf {\bibinfo {volume} {113}},\ \bibinfo
  {pages} {13184} (\bibinfo {year} {2009})}\BibitemShut {NoStop}%
\bibitem [{\citenamefont {Wells}\ \emph {et~al.}(2018)\citenamefont {Wells},
  \citenamefont {Barber}, \citenamefont {Kwizera}, \citenamefont {Mukashyaka},\
  and\ \citenamefont {Chen}}]{Wells2018}%
  \BibitemOpen
  \bibfield  {author} {\bibinfo {author} {\bibfnamefont {T.~A.}\ \bibnamefont
  {Wells}}, \bibinfo {author} {\bibfnamefont {V.~J.}\ \bibnamefont {Barber}},
  \bibinfo {author} {\bibfnamefont {M.~H.}\ \bibnamefont {Kwizera}}, \bibinfo
  {author} {\bibfnamefont {P.}~\bibnamefont {Mukashyaka}}, \ and\ \bibinfo
  {author} {\bibfnamefont {P.~C.}\ \bibnamefont {Chen}},\ }\href {\doibase
  10.1021/acs.jpca.8b08640} {\bibfield  {journal} {\bibinfo  {journal} {J.
  Phys. Chem. A}\ }\textbf {\bibinfo {volume} {122}},\ \bibinfo {pages} {8794}
  (\bibinfo {year} {2018})}\BibitemShut {NoStop}%
\bibitem [{\citenamefont {Kowzan}\ and\ \citenamefont
  {Allison}(2022)}]{Kowzan2022}%
  \BibitemOpen
  \bibfield  {author} {\bibinfo {author} {\bibfnamefont {G.}~\bibnamefont
  {Kowzan}}\ and\ \bibinfo {author} {\bibfnamefont {T.~K.}\ \bibnamefont
  {Allison}},\ }\href {\doibase 10.48550/ARXIV.2206.10488} {\enquote {\bibinfo
  {title} {Theory of rotationally-resolved two-dimensional infrared
  spectroscopy},}\ } (\bibinfo {year} {2022}),\ \Eprint
  {http://arxiv.org/abs/2206.10488} {arXiv:2206.10488 [physics.optics]}
  \BibitemShut {NoStop}%
\bibitem [{\citenamefont {Kowzan}(2022)}]{Kowzan2022rotsim2d}%
  \BibitemOpen
  \bibfield  {author} {\bibinfo {author} {\bibfnamefont {G.}~\bibnamefont
  {Kowzan}},\ }\href {\doibase 10.5281/zenodo.6654257} {\enquote {\bibinfo
  {title} {rotsim2d: {Simulate} {2D} rovibrational spectra of gas-phase
  molecular samples},}\ } (\bibinfo {year} {2022}),\ \bibinfo {note} {{Zenodo}.
  https://doi.org/10.5281/zenodo.6654257}\BibitemShut {NoStop}%
\bibitem [{\citenamefont {Scheurer}\ and\ \citenamefont
  {Mukamel}(2001)}]{Scheurer2001}%
  \BibitemOpen
  \bibfield  {author} {\bibinfo {author} {\bibfnamefont {C.}~\bibnamefont
  {Scheurer}}\ and\ \bibinfo {author} {\bibfnamefont {S.}~\bibnamefont
  {Mukamel}},\ }\href {\doibase 10.1063/1.1391266} {\bibfield  {journal}
  {\bibinfo  {journal} {J. Chem. Phys.}\ }\textbf {\bibinfo {volume} {115}},\
  \bibinfo {pages} {4989} (\bibinfo {year} {2001})}\BibitemShut {NoStop}%
\bibitem [{\citenamefont {Khalil}\ \emph {et~al.}(2003)\citenamefont {Khalil},
  \citenamefont {Demird\"oven},\ and\ \citenamefont {Tokmakoff}}]{Khalil2003}%
  \BibitemOpen
  \bibfield  {author} {\bibinfo {author} {\bibfnamefont {M.}~\bibnamefont
  {Khalil}}, \bibinfo {author} {\bibfnamefont {N.}~\bibnamefont
  {Demird\"oven}}, \ and\ \bibinfo {author} {\bibfnamefont {A.}~\bibnamefont
  {Tokmakoff}},\ }\href {\doibase 10.1103/PhysRevLett.90.047401} {\bibfield
  {journal} {\bibinfo  {journal} {Phys. Rev. Lett.}\ }\textbf {\bibinfo
  {volume} {90}},\ \bibinfo {pages} {047401} (\bibinfo {year}
  {2003})}\BibitemShut {NoStop}%
\bibitem [{\citenamefont {Hamm}\ and\ \citenamefont {Zanni}(2011)}]{Hamm2011a}%
  \BibitemOpen
  \bibfield  {author} {\bibinfo {author} {\bibfnamefont {P.}~\bibnamefont
  {Hamm}}\ and\ \bibinfo {author} {\bibfnamefont {M.}~\bibnamefont {Zanni}},\
  }\href@noop {} {{\selectlanguage {english}\emph {\bibinfo {title} {Concepts
  and methods of {2D} infrared spectroscopy}}}},\ \bibinfo {edition} {1st}\
  ed.\ (\bibinfo  {publisher} {Cambridge University Press},\ \bibinfo {address}
  {Cambridge},\ \bibinfo {year} {2011})\BibitemShut {NoStop}%
\bibitem [{\citenamefont {Gordon}\ \emph {et~al.}(2017)\citenamefont {Gordon},
  \citenamefont {Rothman}, \citenamefont {Hill}, \citenamefont {Kochanov},
  \citenamefont {Tan}, \citenamefont {Bernath}, \citenamefont {Birk},
  \citenamefont {Boudon}, \citenamefont {Campargue}, \citenamefont {Chance},
  \citenamefont {Drouin}, \citenamefont {Flaud}, \citenamefont {Gamache},
  \citenamefont {Hodges}, \citenamefont {Jacquemart}, \citenamefont
  {Perevalov}, \citenamefont {Perrin}, \citenamefont {Shine}, \citenamefont
  {Smith}, \citenamefont {Tennyson}, \citenamefont {Toon}, \citenamefont
  {Tran}, \citenamefont {Tyuterev}, \citenamefont {Barbe}, \citenamefont
  {Cs{\'a}sz{\'a}r}, \citenamefont {Devi}, \citenamefont {Furtenbacher},
  \citenamefont {Harrison}, \citenamefont {Hartmann}, \citenamefont {Jolly},
  \citenamefont {Johnson}, \citenamefont {Karman}, \citenamefont {Kleiner},
  \citenamefont {Kyuberis}, \citenamefont {Loos}, \citenamefont {Lyulin},
  \citenamefont {Massie}, \citenamefont {Mikhailenko}, \citenamefont
  {Moazzen-Ahmadi}, \citenamefont {M{\"u}ller}, \citenamefont {Naumenko},
  \citenamefont {Nikitin}, \citenamefont {Polyansky}, \citenamefont {Rey},
  \citenamefont {Rotger}, \citenamefont {Sharpe}, \citenamefont {Sung},
  \citenamefont {Starikova}, \citenamefont {Tashkun}, \citenamefont {Auwera},
  \citenamefont {Wagner}, \citenamefont {Wilzewski}, \citenamefont {Wcisło},
  \citenamefont {Yu},\ and\ \citenamefont {Zak}}]{Gordon2017}%
  \BibitemOpen
  \bibfield  {author} {\bibinfo {author} {\bibfnamefont {I.}~\bibnamefont
  {Gordon}}, \bibinfo {author} {\bibfnamefont {L.}~\bibnamefont {Rothman}},
  \bibinfo {author} {\bibfnamefont {C.}~\bibnamefont {Hill}}, \bibinfo {author}
  {\bibfnamefont {R.}~\bibnamefont {Kochanov}}, \bibinfo {author}
  {\bibfnamefont {Y.}~\bibnamefont {Tan}}, \bibinfo {author} {\bibfnamefont
  {P.}~\bibnamefont {Bernath}}, \bibinfo {author} {\bibfnamefont
  {M.}~\bibnamefont {Birk}}, \bibinfo {author} {\bibfnamefont {V.}~\bibnamefont
  {Boudon}}, \bibinfo {author} {\bibfnamefont {A.}~\bibnamefont {Campargue}},
  \bibinfo {author} {\bibfnamefont {K.}~\bibnamefont {Chance}}, \bibinfo
  {author} {\bibfnamefont {B.}~\bibnamefont {Drouin}}, \bibinfo {author}
  {\bibfnamefont {J.-M.}\ \bibnamefont {Flaud}}, \bibinfo {author}
  {\bibfnamefont {R.}~\bibnamefont {Gamache}}, \bibinfo {author} {\bibfnamefont
  {J.}~\bibnamefont {Hodges}}, \bibinfo {author} {\bibfnamefont
  {D.}~\bibnamefont {Jacquemart}}, \bibinfo {author} {\bibfnamefont
  {V.}~\bibnamefont {Perevalov}}, \bibinfo {author} {\bibfnamefont
  {A.}~\bibnamefont {Perrin}}, \bibinfo {author} {\bibfnamefont
  {K.}~\bibnamefont {Shine}}, \bibinfo {author} {\bibfnamefont {M.-A.}\
  \bibnamefont {Smith}}, \bibinfo {author} {\bibfnamefont {J.}~\bibnamefont
  {Tennyson}}, \bibinfo {author} {\bibfnamefont {G.}~\bibnamefont {Toon}},
  \bibinfo {author} {\bibfnamefont {H.}~\bibnamefont {Tran}}, \bibinfo {author}
  {\bibfnamefont {V.}~\bibnamefont {Tyuterev}}, \bibinfo {author}
  {\bibfnamefont {A.}~\bibnamefont {Barbe}}, \bibinfo {author} {\bibfnamefont
  {A.}~\bibnamefont {Cs{\'a}sz{\'a}r}}, \bibinfo {author} {\bibfnamefont
  {V.}~\bibnamefont {Devi}}, \bibinfo {author} {\bibfnamefont {T.}~\bibnamefont
  {Furtenbacher}}, \bibinfo {author} {\bibfnamefont {J.}~\bibnamefont
  {Harrison}}, \bibinfo {author} {\bibfnamefont {J.-M.}\ \bibnamefont
  {Hartmann}}, \bibinfo {author} {\bibfnamefont {A.}~\bibnamefont {Jolly}},
  \bibinfo {author} {\bibfnamefont {T.}~\bibnamefont {Johnson}}, \bibinfo
  {author} {\bibfnamefont {T.}~\bibnamefont {Karman}}, \bibinfo {author}
  {\bibfnamefont {I.}~\bibnamefont {Kleiner}}, \bibinfo {author} {\bibfnamefont
  {A.}~\bibnamefont {Kyuberis}}, \bibinfo {author} {\bibfnamefont
  {J.}~\bibnamefont {Loos}}, \bibinfo {author} {\bibfnamefont {O.}~\bibnamefont
  {Lyulin}}, \bibinfo {author} {\bibfnamefont {S.}~\bibnamefont {Massie}},
  \bibinfo {author} {\bibfnamefont {S.}~\bibnamefont {Mikhailenko}}, \bibinfo
  {author} {\bibfnamefont {N.}~\bibnamefont {Moazzen-Ahmadi}}, \bibinfo
  {author} {\bibfnamefont {H.}~\bibnamefont {M{\"u}ller}}, \bibinfo {author}
  {\bibfnamefont {O.}~\bibnamefont {Naumenko}}, \bibinfo {author}
  {\bibfnamefont {A.}~\bibnamefont {Nikitin}}, \bibinfo {author} {\bibfnamefont
  {O.}~\bibnamefont {Polyansky}}, \bibinfo {author} {\bibfnamefont
  {M.}~\bibnamefont {Rey}}, \bibinfo {author} {\bibfnamefont {M.}~\bibnamefont
  {Rotger}}, \bibinfo {author} {\bibfnamefont {S.}~\bibnamefont {Sharpe}},
  \bibinfo {author} {\bibfnamefont {K.}~\bibnamefont {Sung}}, \bibinfo {author}
  {\bibfnamefont {E.}~\bibnamefont {Starikova}}, \bibinfo {author}
  {\bibfnamefont {S.}~\bibnamefont {Tashkun}}, \bibinfo {author} {\bibfnamefont
  {J.~V.}\ \bibnamefont {Auwera}}, \bibinfo {author} {\bibfnamefont
  {G.}~\bibnamefont {Wagner}}, \bibinfo {author} {\bibfnamefont
  {J.}~\bibnamefont {Wilzewski}}, \bibinfo {author} {\bibfnamefont
  {P.}~\bibnamefont {Wcisło}}, \bibinfo {author} {\bibfnamefont
  {S.}~\bibnamefont {Yu}}, \ and\ \bibinfo {author} {\bibfnamefont
  {E.}~\bibnamefont {Zak}},\ }\href {\doibase 10.1016/j.jqsrt.2017.06.038}
  {\bibfield  {journal} {\bibinfo  {journal} {J. Quant. Spectrosc. Radiat.
  Transfer}\ }\textbf {\bibinfo {volume} {203}},\ \bibinfo {pages} {3}
  (\bibinfo {year} {2017})}\BibitemShut {NoStop}%
\bibitem [{\citenamefont {Nikitin}\ \emph {et~al.}(2005)\citenamefont
  {Nikitin}, \citenamefont {Champion},\ and\ \citenamefont
  {B{\"u}rger}}]{Nikitin2005}%
  \BibitemOpen
  \bibfield  {author} {\bibinfo {author} {\bibfnamefont {A.}~\bibnamefont
  {Nikitin}}, \bibinfo {author} {\bibfnamefont {J.}~\bibnamefont {Champion}}, \
  and\ \bibinfo {author} {\bibfnamefont {H.}~\bibnamefont {B{\"u}rger}},\
  }\href {\doibase 10.1016/j.jms.2004.11.012} {\bibfield  {journal} {\bibinfo
  {journal} {J. Mol. Spectrosc.}\ }\textbf {\bibinfo {volume} {230}},\ \bibinfo
  {pages} {174} (\bibinfo {year} {2005})}\BibitemShut {NoStop}%
\bibitem [{\citenamefont {Zare}(1991)}]{Zare1991}%
  \BibitemOpen
  \bibfield  {author} {\bibinfo {author} {\bibfnamefont {R.~N.}\ \bibnamefont
  {Zare}},\ }\href@noop {} {\emph {\bibinfo {title} {Angular Momentum:
  Understanding Spatial Aspects in Chemistry and Physics}}},\ \bibinfo
  {edition} {1st}\ ed.\ (\bibinfo  {publisher} {Wiley-Interscience},\ \bibinfo
  {year} {1991})\BibitemShut {NoStop}%
\bibitem [{\citenamefont {Blum}(2012)}]{Blum2012}%
  \BibitemOpen
  \bibfield  {author} {\bibinfo {author} {\bibfnamefont {K.}~\bibnamefont
  {Blum}},\ }\href {\doibase 10.1007/978-3-642-20561-3} {\emph {\bibinfo
  {title} {Density Matrix Theory and Applications}}},\ \bibinfo {edition}
  {3rd}\ ed.,\ Springer Series on Atomic, Optical, and Plasma Physics\
  (\bibinfo  {publisher} {Springer},\ \bibinfo {year} {2012})\BibitemShut
  {NoStop}%
\bibitem [{\citenamefont {Weiner}(2009)}]{Weiner2009}%
  \BibitemOpen
  \bibfield  {author} {\bibinfo {author} {\bibfnamefont {A.}~\bibnamefont
  {Weiner}},\ }\href@noop {} {\emph {\bibinfo {title} {Ultrafast optics}}}\
  (\bibinfo  {publisher} {Wiley},\ \bibinfo {address} {Hoboken, N.J},\ \bibinfo
  {year} {2009})\BibitemShut {NoStop}%
\bibitem [{\citenamefont {Benner}\ \emph {et~al.}(1995)\citenamefont {Benner},
  \citenamefont {Rinsland}, \citenamefont {Devi}, \citenamefont {Smith},\ and\
  \citenamefont {Atkins}}]{Benner1995}%
  \BibitemOpen
  \bibfield  {author} {\bibinfo {author} {\bibfnamefont {D.~C.}\ \bibnamefont
  {Benner}}, \bibinfo {author} {\bibfnamefont {C.~P.}\ \bibnamefont
  {Rinsland}}, \bibinfo {author} {\bibfnamefont {V.~M.}\ \bibnamefont {Devi}},
  \bibinfo {author} {\bibfnamefont {M.~A.~H.}\ \bibnamefont {Smith}}, \ and\
  \bibinfo {author} {\bibfnamefont {D.}~\bibnamefont {Atkins}},\ }\href
  {\doibase 10.1016/0022-4073(95)00015-D} {\bibfield  {journal} {\bibinfo
  {journal} {J. Quant. Spectrosc. Radiat. Transfer}\ }\textbf {\bibinfo
  {volume} {53}},\ \bibinfo {pages} {705} (\bibinfo {year} {1995})}\BibitemShut
  {NoStop}%
\bibitem [{\citenamefont {Hartmann}\ \emph {et~al.}(2008)\citenamefont
  {Hartmann}, \citenamefont {Boulet},\ and\ \citenamefont
  {Robert}}]{Hartmann2008}%
  \BibitemOpen
  \bibfield  {author} {\bibinfo {author} {\bibfnamefont {J.-M.}\ \bibnamefont
  {Hartmann}}, \bibinfo {author} {\bibfnamefont {C.}~\bibnamefont {Boulet}}, \
  and\ \bibinfo {author} {\bibfnamefont {D.}~\bibnamefont {Robert}},\
  }\href@noop {} {\emph {\bibinfo {title} {Collisional effects on molecular
  spectra. Laboratory experiments and models, consequences for applications}}}\
  (\bibinfo  {publisher} {Elsevier},\ \bibinfo {address} {Amsterdam},\ \bibinfo
  {year} {2008})\BibitemShut {NoStop}%
\bibitem [{\citenamefont {{van Stokkum}}\ \emph {et~al.}(2004)\citenamefont
  {{van Stokkum}}, \citenamefont {Larsen},\ and\ \citenamefont {{van
  Grondelle}}}]{Stokkum2004}%
  \BibitemOpen
  \bibfield  {author} {\bibinfo {author} {\bibfnamefont {I.~H.}\ \bibnamefont
  {{van Stokkum}}}, \bibinfo {author} {\bibfnamefont {D.~S.}\ \bibnamefont
  {Larsen}}, \ and\ \bibinfo {author} {\bibfnamefont {R.}~\bibnamefont {{van
  Grondelle}}},\ }\href {\doibase https://doi.org/10.1016/j.bbabio.2004.04.011}
  {\bibfield  {journal} {\bibinfo  {journal} {Biochim. Biophys. Acta,
  Bioenerg.}\ }\textbf {\bibinfo {volume} {1657}},\ \bibinfo {pages} {82 }
  (\bibinfo {year} {2004})}\BibitemShut {NoStop}%
\bibitem [{\citenamefont {Slavov}\ \emph {et~al.}(2015)\citenamefont {Slavov},
  \citenamefont {Hartmann},\ and\ \citenamefont {Wachtveitl}}]{Slavov2015}%
  \BibitemOpen
  \bibfield  {author} {\bibinfo {author} {\bibfnamefont {C.}~\bibnamefont
  {Slavov}}, \bibinfo {author} {\bibfnamefont {H.}~\bibnamefont {Hartmann}}, \
  and\ \bibinfo {author} {\bibfnamefont {J.}~\bibnamefont {Wachtveitl}},\
  }\href {\doibase 10.1021/ac504348h} {\bibfield  {journal} {\bibinfo
  {journal} {Anal. Chem.}\ }\textbf {\bibinfo {volume} {87}},\ \bibinfo {pages}
  {2328} (\bibinfo {year} {2015})}\BibitemShut {NoStop}%
\bibitem [{\citenamefont {Chew}\ \emph {et~al.}(2013)\citenamefont {Chew},
  \citenamefont {Nemchick},\ and\ \citenamefont {Vaccaro}}]{Chew2013}%
  \BibitemOpen
  \bibfield  {author} {\bibinfo {author} {\bibfnamefont {K.}~\bibnamefont
  {Chew}}, \bibinfo {author} {\bibfnamefont {D.~J.}\ \bibnamefont {Nemchick}},
  \ and\ \bibinfo {author} {\bibfnamefont {P.~H.}\ \bibnamefont {Vaccaro}},\
  }\href {\doibase 10.1021/jp400160z} {\bibfield  {journal} {\bibinfo
  {journal} {J. Phys. Chem. A}\ }\textbf {\bibinfo {volume} {117}},\ \bibinfo
  {pages} {6126} (\bibinfo {year} {2013})}\BibitemShut {NoStop}%
\bibitem [{\citenamefont {Helbing}\ and\ \citenamefont
  {Hamm}(2011)}]{Helbing2010}%
  \BibitemOpen
  \bibfield  {author} {\bibinfo {author} {\bibfnamefont {J.}~\bibnamefont
  {Helbing}}\ and\ \bibinfo {author} {\bibfnamefont {P.}~\bibnamefont {Hamm}},\
  }\href {\doibase 10.1364/josab.28.000171} {\bibfield  {journal} {\bibinfo
  {journal} {J. Opt. Soc. Am. B}\ }\textbf {\bibinfo {volume} {28}},\ \bibinfo
  {pages} {171} (\bibinfo {year} {2011})}\BibitemShut {NoStop}%
\bibitem [{\citenamefont {Chen}\ \emph {et~al.}(2009)\citenamefont {Chen},
  \citenamefont {Settersten},\ and\ \citenamefont {Kouzov}}]{Chen2009}%
  \BibitemOpen
  \bibfield  {author} {\bibinfo {author} {\bibfnamefont {X.}~\bibnamefont
  {Chen}}, \bibinfo {author} {\bibfnamefont {T.~B.}\ \bibnamefont
  {Settersten}}, \ and\ \bibinfo {author} {\bibfnamefont {A.~P.}\ \bibnamefont
  {Kouzov}},\ }\href {\doibase 10.1002/jrs.2229} {\bibfield  {journal}
  {\bibinfo  {journal} {J. Raman Spectrosc.}\ }\textbf {\bibinfo {volume}
  {40}},\ \bibinfo {pages} {847} (\bibinfo {year} {2009})}\BibitemShut
  {NoStop}%
\bibitem [{\citenamefont {Pack}\ \emph {et~al.}(2019)\citenamefont {Pack},
  \citenamefont {Rotondaro}, \citenamefont {Shah}, \citenamefont {Mandal},
  \citenamefont {Erramilli},\ and\ \citenamefont {Ziegler}}]{Pack2019}%
  \BibitemOpen
  \bibfield  {author} {\bibinfo {author} {\bibfnamefont {G.~N.}\ \bibnamefont
  {Pack}}, \bibinfo {author} {\bibfnamefont {M.~C.}\ \bibnamefont {Rotondaro}},
  \bibinfo {author} {\bibfnamefont {P.~P.}\ \bibnamefont {Shah}}, \bibinfo
  {author} {\bibfnamefont {A.}~\bibnamefont {Mandal}}, \bibinfo {author}
  {\bibfnamefont {S.}~\bibnamefont {Erramilli}}, \ and\ \bibinfo {author}
  {\bibfnamefont {L.~D.}\ \bibnamefont {Ziegler}},\ }\href {\doibase
  10.1039/c9cp04101j} {\bibfield  {journal} {\bibinfo  {journal} {Phys. Chem.
  Chem. Phys.}\ }\textbf {\bibinfo {volume} {21}},\ \bibinfo {pages} {21249}
  (\bibinfo {year} {2019})}\BibitemShut {NoStop}%
\bibitem [{\citenamefont {Henderson}\ \emph {et~al.}(2018)\citenamefont
  {Henderson}, \citenamefont {Khodabakhsh}, \citenamefont {Mets{\"a}l{\"a}},
  \citenamefont {Ventrillard}, \citenamefont {Schmidt}, \citenamefont
  {Romanini}, \citenamefont {Ritchie}, \citenamefont {te~Lintel~Hekkert},
  \citenamefont {Briot}, \citenamefont {Risby}, \citenamefont {Marczin},
  \citenamefont {Harren},\ and\ \citenamefont {Cristescu}}]{Henderson2018a}%
  \BibitemOpen
  \bibfield  {author} {\bibinfo {author} {\bibfnamefont {B.}~\bibnamefont
  {Henderson}}, \bibinfo {author} {\bibfnamefont {A.}~\bibnamefont
  {Khodabakhsh}}, \bibinfo {author} {\bibfnamefont {M.}~\bibnamefont
  {Mets{\"a}l{\"a}}}, \bibinfo {author} {\bibfnamefont {I.}~\bibnamefont
  {Ventrillard}}, \bibinfo {author} {\bibfnamefont {F.~M.}\ \bibnamefont
  {Schmidt}}, \bibinfo {author} {\bibfnamefont {D.}~\bibnamefont {Romanini}},
  \bibinfo {author} {\bibfnamefont {G.~A.~D.}\ \bibnamefont {Ritchie}},
  \bibinfo {author} {\bibfnamefont {S.}~\bibnamefont {te~Lintel~Hekkert}},
  \bibinfo {author} {\bibfnamefont {R.}~\bibnamefont {Briot}}, \bibinfo
  {author} {\bibfnamefont {T.}~\bibnamefont {Risby}}, \bibinfo {author}
  {\bibfnamefont {N.}~\bibnamefont {Marczin}}, \bibinfo {author} {\bibfnamefont
  {F.~J.~M.}\ \bibnamefont {Harren}}, \ and\ \bibinfo {author} {\bibfnamefont
  {S.~M.}\ \bibnamefont {Cristescu}},\ }\href {\doibase
  10.1007/s00340-018-7030-x} {\bibfield  {journal} {\bibinfo  {journal} {Appl.
  Phys. B: Lasers Opt.}\ }\textbf {\bibinfo {volume} {124}},\ \bibinfo {pages}
  {161} (\bibinfo {year} {2018})}\BibitemShut {NoStop}%
\bibitem [{\citenamefont {Liang}\ \emph {et~al.}(2021)\citenamefont {Liang},
  \citenamefont {Chan}, \citenamefont {Changala}, \citenamefont {Nesbitt},
  \citenamefont {Ye},\ and\ \citenamefont {Toscano}}]{Liang2021a}%
  \BibitemOpen
  \bibfield  {author} {\bibinfo {author} {\bibfnamefont {Q.}~\bibnamefont
  {Liang}}, \bibinfo {author} {\bibfnamefont {Y.-C.}\ \bibnamefont {Chan}},
  \bibinfo {author} {\bibfnamefont {P.~B.}\ \bibnamefont {Changala}}, \bibinfo
  {author} {\bibfnamefont {D.~J.}\ \bibnamefont {Nesbitt}}, \bibinfo {author}
  {\bibfnamefont {J.}~\bibnamefont {Ye}}, \ and\ \bibinfo {author}
  {\bibfnamefont {J.}~\bibnamefont {Toscano}},\ }\href {\doibase
  10.1073/pnas.2105063118} {\bibfield  {journal} {\bibinfo  {journal} {Proc.
  Natl. Acad. Sci.}\ }\textbf {\bibinfo {volume} {118}},\ \bibinfo {pages}
  {e2105063118} (\bibinfo {year} {2021})}\BibitemShut {NoStop}%
\bibitem [{\citenamefont {Chen}\ and\ \citenamefont
  {Settersten}(2007)}]{Chen2007}%
  \BibitemOpen
  \bibfield  {author} {\bibinfo {author} {\bibfnamefont {X.}~\bibnamefont
  {Chen}}\ and\ \bibinfo {author} {\bibfnamefont {T.~B.}\ \bibnamefont
  {Settersten}},\ }\href {\doibase 10.1364/ao.46.003911} {\bibfield  {journal}
  {\bibinfo  {journal} {Appl. Opt.}\ }\textbf {\bibinfo {volume} {46}},\
  \bibinfo {pages} {3911} (\bibinfo {year} {2007})}\BibitemShut {NoStop}%
\bibitem [{\citenamefont {Patnaik}\ \emph {et~al.}(2017)\citenamefont
  {Patnaik}, \citenamefont {Adamovich}, \citenamefont {Gord},\ and\
  \citenamefont {Roy}}]{Patnaik2017}%
  \BibitemOpen
  \bibfield  {author} {\bibinfo {author} {\bibfnamefont {A.~K.}\ \bibnamefont
  {Patnaik}}, \bibinfo {author} {\bibfnamefont {I.}~\bibnamefont {Adamovich}},
  \bibinfo {author} {\bibfnamefont {J.~R.}\ \bibnamefont {Gord}}, \ and\
  \bibinfo {author} {\bibfnamefont {S.}~\bibnamefont {Roy}},\ }\href {\doibase
  10.1088/1361-6595/aa8578} {\bibfield  {journal} {\bibinfo  {journal} {Plasma
  Sources Sci. Technol.}\ }\textbf {\bibinfo {volume} {26}},\ \bibinfo {pages}
  {103001} (\bibinfo {year} {2017})}\BibitemShut {NoStop}%
\end{thebibliography}
\end{document}